\documentclass[twocolumn,showpacs,aps,prb,superscriptaddress,floatfix]{revtex4}

\usepackage{graphicx}
\usepackage{dcolumn}
\usepackage{bm}
\begin{document}
\preprint{}

\title{Spin-Orbit Coupling in Iridium-Based 5d Compounds Probed by X-ray Absorption Spectroscopy}

\author{J.P. Clancy}
\affiliation{Department of Physics, University of Toronto, Toronto, Ontario, M5S 1A7, Canada}

\author{N. Chen}
\affiliation{Canadian Light Source, Saskatoon, Saskatchewan, S7N 0X4, Canada}

\author{C.Y. Kim}
\affiliation{Canadian Light Source, Saskatoon, Saskatchewan, S7N 0X4, Canada}

\author{W.F. Chen}
\affiliation{Canadian Light Source, Saskatoon, Saskatchewan, S7N 0X4, Canada}

\author{K.W. Plumb}
\affiliation{Department of Physics, University of Toronto, Toronto, Ontario, M5S 1A7, Canada}

\author{B.C. Jeon}
\affiliation{ReCFI, Department of Physics and Astronomy, Seoul National University, Seoul 151-747, Korea}

\author{T.W. Noh}
\affiliation{ReCFI, Department of Physics and Astronomy, Seoul National University, Seoul 151-747, Korea}

\author{Young-June Kim}
\affiliation{Department of Physics, University of Toronto, Toronto, Ontario, M5S 1A7, Canada}

\begin{abstract}
We have performed x-ray absorption spectroscopy (XAS) measurements on a series of Ir-based 5d transition metal compounds, including Ir, IrCl$_3$, IrO$_2$, Na$_2$IrO$_3$, Sr$_2$IrO$_4$, and Y$_2$Ir$_2$O$_7$.  By comparing the intensity of the ``white-line'' features observed at the Ir L$_2$ and L$_3$ absorption edges, it is possible to extract valuable information about the strength of the spin-orbit coupling in these systems.  We observe remarkably large, non-statistical branching ratios in all Ir compounds studied, with little or no dependence on chemical composition, crystal structure, or electronic state.  This result confirms the presence of strong spin-orbit coupling effects in novel iridates such as Sr$_2$IrO$_4$, Na$_2$IrO$_3$, and Y$_2$Ir$_2$O$_7$, and suggests that even simple Ir-based compounds such as IrO$_2$ and IrCl$_3$ may warrant further study.  In contrast, XAS measurements on Re-based 5d compounds, such as Re, ReO$_2$, ReO$_3$, and Ba$_2$FeReO$_6$, reveal statistical branching ratios and negligible spin-orbit coupling effects.

\end{abstract}
\maketitle

\section{Introduction}
In recent years there has been growing interest in the physics of materials based on 5d transition metal elements\cite{Kim_PRL_2008, Kim_Science_2009, Moon_PRL_2008, Cao_PRB_1998, Crawford_PRB_1994, Huang_JSSC_1994, Ge_PRB_2011, Chikara_PRB_2009, Kim_PRL_2012, Cao_PRB_2002, Singh_PRB_2010, Liu_PRB_2011, Singh_PRL_2012, Choi_PRL_2012, Ye_arXiv_2012, Taira_JPCM_2001, Shitade_PRL_2009, Jackeli_PRL_2009, Chaloupka_PRL_2010, Nakatsuji_PRL_2006, Machida_Nature_2010, Pesin_NP_2010, Yang_PRB_2010, Wan_PRB_2011, Witczak-Krempa_PRB_2012}.  Due to the broad, spatially extended nature of 5d electronic wavefunctions, these materials tend to exhibit wide electronic bands, with relatively weak electronic correlations (U $\sim$ 0.5 - 3 eV) and strong crystal field effects (CF $\sim$ 1 - 5 eV).  Furthermore, because of the large atomic mass associated with 5d elements, these materials also tend to exhibit strong relativistic spin-orbit coupling effects (SO $\sim$ 0.1 - 1 eV).  The close correspondence of these three energy scales (U $\approx$ CF $\approx$ SO) presents an extremely promising opportunity for the study of novel physics arising from competing spin, orbital, charge, and lattice degrees of freedom.

One family of 5d compounds which have begun to attract particular experimental and theoretical attention are the iridium-based transition metal oxides, or iridates\cite{Kim_PRL_2008, Kim_Science_2009, Moon_PRL_2008, Cao_PRB_1998, Crawford_PRB_1994, Huang_JSSC_1994, Ge_PRB_2011, Chikara_PRB_2009, Kim_PRL_2012, Cao_PRB_2002, Singh_PRB_2010, Liu_PRB_2011, Singh_PRL_2012, Choi_PRL_2012, Ye_arXiv_2012, Taira_JPCM_2001, Shitade_PRL_2009, Jackeli_PRL_2009, Chaloupka_PRL_2010, Nakatsuji_PRL_2006, Machida_Nature_2010, Pesin_NP_2010, Yang_PRB_2010, Wan_PRB_2011, Witczak-Krempa_PRB_2012}.  The majority of these compounds are based on Ir$^{4+}$ ions, which display a 5d$^5$ electronic configuration.  One of the fundamental puzzles surrounding these materials is that in spite of the fact that simple band theory arguments suggest they should be metallic, a surprising number of iridates appear to be insulators\cite{Moon_PRL_2008, Cao_PRB_1998, Cao_PRB_2002, Singh_PRB_2010}.  This result is particularly remarkable since the relatively weak 5d electron-electron correlations are too small to drive a conventional Mott insulator transition (as in many 3d transition metal oxide systems).  Instead, this phenomena has been attributed to the presence of much stronger 5d spin-orbit coupling effects, which can be combined with the modest electronic correlations to produce a novel ``spin-orbital Mott insulator'' with a j$_{eff}$ = 1/2 ground state\cite{Kim_PRL_2008, Kim_Science_2009}.

Among the iridates that have attracted the most significant recent attention are Sr$_2$IrO$_4$\cite{Kim_PRL_2008, Kim_Science_2009, Moon_PRL_2008, Cao_PRB_1998, Crawford_PRB_1994, Huang_JSSC_1994, Ge_PRB_2011, Chikara_PRB_2009, Kim_PRL_2012, Jackeli_PRL_2009}, Na$_2$IrO$_3$\cite{Singh_PRB_2010, Liu_PRB_2011, Singh_PRL_2012, Choi_PRL_2012, Ye_arXiv_2012, Shitade_PRL_2009, Chaloupka_PRL_2010}, and Y$_2$Ir$_2$O$_7$\cite{Taira_JPCM_2001, Pesin_NP_2010, Yang_PRB_2010, Wan_PRB_2011, Witczak-Krempa_PRB_2012}.  Sr$_2$IrO$_4$ is a layered perovskite compound, with a structure similar to the parent compound of the high T$_c$ cuprates, La$_2$CuO$_4$\cite{Crawford_PRB_1994, Huang_JSSC_1994}.  This compound develops canted antiferromagnetic order below T$_N$ $\sim$ 240 K\cite{Kim_Science_2009}, and has been reported to display unusually large magnetoelastic\cite{Ge_PRB_2011} and magnetoelectric effects\cite{Chikara_PRB_2009}.  More importantly, however, Sr$_2$IrO$_4$ is notable for being the first proposed experimental realization of a j$_{eff}$ = 1/2 spin-orbital Mott insulator\cite{Kim_PRL_2008, Kim_Science_2009}.  Na$_2$IrO$_3$ is a honeycomb lattice material, which has attracted considerable interest due to its potential relevance to the Kitaev-Heisenberg model\cite{Chaloupka_PRL_2010, Liu_PRB_2011, Singh_PRL_2012, Choi_PRL_2012, Ye_arXiv_2012}.  This model, which describes a system of S = 1/2 moments on a honeycomb lattice with highly anisotropic exchange interactions, has been proposed to exhibit novel spin liquid and topologically ordered states\cite{Chaloupka_PRL_2010}.  Y$_2$Ir$_2$O$_7$ is a pyrochlore iridate, with a structure composed of Ir ions distributed on a network of corner-sharing tetrahedra.  This crystal structure gives rise to a natural tendency towards geometric frustration, and members of the R$_2$Ir$_2$O$_7$ series (R = lanthanide) have been associated with exotic spin liquid and spin ice states\cite{Nakatsuji_PRL_2006, Machida_Nature_2010}.  Y$_2$Ir$_2$O$_7$, in particular, has also been put forward as a potential topological insulator or topological (Weyl) semi-metal\cite{Pesin_NP_2010, Yang_PRB_2010, Wan_PRB_2011, Witczak-Krempa_PRB_2012}.  Although the detailed physics of these three systems vary significantly, in all three cases it is the presence of strong 5d spin-orbit coupling effects that is believed to drive the development of unconventional electronic and magnetic ground states.

Given the apparent importance of spin-orbit coupling effects in the iridates, it is natural to attempt to quantify the strength of the spin-orbit interactions in these compounds.  One method of accomplishing this is provided by the technique of X-ray Absorption Spectroscopy (XAS).  This method was initially proposed by van der Laan and Thole\cite{vanderLaan_PRL_1988, Thole_PRB_1988, Thole_PRA_1988}, who recognized that the intensity ratio of the ``white-line'' features observed in XAS measurements at the L$_2$ and L$_3$ absorption edges is directly proportional to the expectation value of the spin-orbit operator $\langle$L$\cdot$S$\rangle$.  As a result, XAS provides a direct probe of spin-orbit interactions, which is complementary to other techniques such as magnetic susceptibility, electron paramagnetic resonance, and Mossbauer spectroscopy (which probe spin-orbit coupling through the value of the Lande g-factor) or optical spectroscopy (which probes spin-orbit coupling through the observation of transitions forbidden by spin-only selection rules).  This method was most recently employed by Laguna-Marco et al to investigate spin-orbit coupling in Ba$_{1-x}$Sr$_x$IrO$_3$ (x = 0, 0.06, 0.12)\cite{Laguna-Marco_PRL_2010}.  This study revealed a surprisingly large L$_3$/L$_2$ branching ratio, providing confirmation of the unusually strong spin-orbit interactions believed to be present in that compound.

In this paper, we report x-ray absorption spectroscopy (XAS) measurements on a series of iridium-based 5d compounds.  In particular, we have carried out a comparative study of several topical iridates (Sr$_2$IrO$_4$, Na$_2$IrO$_3$, and Y$_2$Ir$_2$O$_7$) and several standard reference samples such as Ir, IrCl$_3$, and IrO$_2$.  Our measurements reveal anomalously large branching ratios in all Ir-based compounds studied, and even (to a lesser extent) in elemental iridium itself.  On the basis of these results, we propose that unusually strong spin-orbit coupling effects may be a common feature of all iridates, or at least all those possessing an octahedral local crystal field environment.  To provide a point of comparison, similar measurements were also performed on a series of rhenium-based 5d compounds, including Re, ReO$_2$, Ba$_2$FeReO$_6$, and ReO$_3$.  These compounds present a striking contrast to the iridates, displaying small, statistical branching ratios, which are consistent with negligible spin-orbit coupling effects.

\section{Experimental Details}

X-ray absorption measurements were performed on a series of six Iridium-based compounds, including Ir, IrCl$_3$, IrO$_2$, Na$_2$IrO$_3$, Sr$_2$IrO$_4$, and Y$_2$Ir$_2$O$_7$.  These compounds encompass a wide range of physical properties, displaying a variety of electronic configurations (5d$^7$, 5d$^6$, and 5d$^5$), ionization states (Ir$^0$, Ir$^{3+}$, Ir$^{4+}$), transport properties (metallic, insulating), and crystal structures (cubic, tetragonal, orthorhombic, and monoclinic).  However, one common feature shared by all of these compounds (with the exception of elemental Ir) is the presence of an octahedral local crystal field environment, wherein each Ir ion is surrounded by six nearest neighbor O or Cl ions.  A more detailed description of the physical properties associated with these samples is provided in Table I.  In order to provide a point of comparison, XAS measurements were also performed on a series of Rhenium-based samples, including Re, ReO$_2$, Ba$_2$FeReO$_6$, and ReO$_3$.  As in the case of the Ir-based samples, these Re-based compounds display a variety of different valence states (Re$^0$ [5d$^5$], Re$^{4+}$ [5d$^3$], Re$^{5+}$ [5d$^2$], Re$^{6+}$ [5d$^1$]) and crystal structures, although all samples exhibit octahedral local coordination.  Measurements were carried out on powder samples obtained from commercial suppliers (Ir, IrCl$_3$, IrO$_2$, Re, ReO$_2$, ReO$_3$) or prepared using conventional solid state reaction methods (Sr$_2$IrO$_4$, Na$_2$IrO$_3$, Y$_2$Ir$_2$O$_7$, Ba$_2$FeReO$_6$)\cite{Singh_PRB_2010, Kim_Science_2009, Taira_JPCM_2001, Jeon_JPCM_2010}.  The purity of the commercial samples was quoted as 99.7\% (ReO$_2$), 99.9\% (ReO$_3$), 99.99\% (Ir, IrCl$_3$, IrO$_2$), and 99.999\% (Re) or higher.  Thin, plate-like, powder samples were prepared on sheets of Kapton tape, with two to three layers of sample used to minimize ``pin-holing'' effects and produce an average sample thickness of 5 - 10 $\mu$m (or $\sim$ 2 absorption lengths).  

\begin{table*}
\begin{tabular}{| c | m{1.7cm} | m{1.7cm} | c | m{1.7cm} | m{2.4cm} | c |}
    \hline
    Compound & \centering{Electronic State} & \centering{Electronic Properties} & Crystal Structure & \centering{Ir Site Symmetry} & \centering{Ir-O Distance (avg, min)} & {Ir-Ir Distance (min)} \\ \hline \hline
		Ir & \centering{5d$^7$ (Ir)} & \centering{Metallic} & \centering{Cubic (\it{Fm$\bar{3}$m})} & \centering{m$\bar{3}$m} & \centering{---} & 2.715{\AA} \\ \hline
    IrCl$_3$ & \centering{5d$^6$ (Ir$^{3+}$)} & \centering{Insulating} & \centering{Monoclinic (\it{C2/m})} & \centering{2} & \centering{---} & 3.457{\AA} \\ \hline
    IrO$_2$ & \centering{5d$^5$ (Ir$^{4+}$)} & \centering{Metallic} & \centering{Orthorhombic (\it{P4$_2$/mnm})} & \centering{mmm} & \centering{1.986{\AA}, 1.960{\AA}} & 3.159{\AA} \\ \hline
    Na$_2$IrO$_3$ & \centering{5d$^5$ (Ir$^{4+}$)} & \centering{Insulating} & \centering{Monoclinic (\it{C2/m})} & \centering{2} & \centering{2.074{\AA}, 2.060{\AA}} & 3.138{\AA} \\ \hline
    Sr$_2$IrO$_4$ & \centering{5d$^5$ (Ir$^{4+}$)} & \centering{Insulating} & \centering{Tetragonal (\it{I4$_1$/acd})} & \centering{$\bar{1}$} & \centering{2.006{\AA}, 1.980{\AA}} & 3.878{\AA} \\ \hline 
    Y$_2$Ir$_2$O$_7$ & \centering{5d$^5$ (Ir$^{4+}$)} & \centering{Metallic*} & \centering{Cubic (\it{Fd$\bar{3}$m})} & \centering{.$\bar{3}$m} & \centering{2.015{\AA}, 2.015{\AA}} & 3.599{\AA} \\ \hline \hline
    Compound & \centering{Electronic State} & \centering{Electronic Properties} & Crystal Structure & \centering{Re Site Symmetry} & \centering{Re-O Distance (avg, min)} & Re-Re Distance (min) \\ \hline \hline
    Re & \centering{5d$^5$ (Re)} & \centering{Metallic} & \centering{Hexagonal (\it{P6$_3$/mmc})} & \centering{$\bar{6}$m2} & \centering{---} & 2.761{\AA} \\ \hline
    ReO$_2$ & \centering{5d$^3$ (Re$^{4+}$)} & \centering{Metallic} & \centering{Orthorhombic (\it{Pbcn})} & \centering{.2.} & \centering{1.997{\AA}, 1.941{\AA}} & 2.614{\AA} \\ \hline
    Ba$_2$FeReO$_6$ & \centering{5d$^2$ (Re$^{5+}$)} & \centering{Metallic} & \centering{Cubic (\it{Fm$\bar{3}$m})} & \centering{m$\bar{3}$m} & \centering{1.932{\AA}, 1.932{\AA}} & 5.717{\AA} \\ \hline
    ReO$_3$ & \centering{5d$^1$ (Re$^{6+}$)} & \centering{Metallic} & \centering{Cubic (\it{Pm$\bar{3}$m})} & \centering{m$\bar{3}$m} & \centering{1.867{\AA}, 1.867{\AA}} & 3.734{\AA} \\ \hline
\end{tabular}
\caption{Basic physical properties of the Ir and Re-based compounds studied in this experiment.  Crystallographic data has been obtained from References 5, 6, 15, 16, 21, 30, and 49-55.  *Note that Y$_2$Ir$_2$O$_7$ undergoes a metal-to-insulator transition at T$_{MI}$ $\sim$ 150 K\cite{Taira_JPCM_2001}, but at room temperature it exhibits weakly metallic properties. }
\end{table*}

X-ray absorption measurements were carried out using the Hard X-ray MicroAnalysis (HXMA) beamline 06ID-1 at the Canadian Light Source (CLS).  Measurements were collected at both the L$_2$ (2p$_{1/2}$ $\rightarrow$ 5d) and L$_3$ (2p$_{3/2}$ $\rightarrow$ 5d) absorption edges, which occur at energies of 12.824 keV and 11.215 keV (Ir) and 11.959 keV and 10.535 keV (Re), respectively.  The energy of the incident x-ray beam was selected using a Si(111) monochromator, with higher harmonic contributions suppressed by a combination of Rh-coated mirrors and a 50\% detuning of the undulator.  XAS measurements were performed in transmission geometry, providing a direct measure of the linear x-ray attenuation coefficient, $\mu$(E).  $\mu$(E) is given by the ratio of the incident intensity (I$_0$) and the transmitted intensity (I$_1$) of the x-ray beam through the sample.  A reference sample, consisting of elemental Ir or Re, was mounted after the sample to allow for energy calibration.  All data was collected at room temperature.

\section{Results and Discussion}

X-ray absorption spectra collected at the Ir L$_2$ and L$_3$ absorption edges are provided in Figure 1 for a series of Ir-based compounds.  Each of these spectra can be decomposed into three main features: (i) a sharp, atomic-like ``white-line'' feature that corresponds to 2p $\rightarrow$ 5d electronic transitions, (ii) a step-like edge feature which is associated with 2p $\rightarrow$ continuum electronic excitations, and (iii) a series of smaller ``fine structure'' oscillations that arise from the backscattering of photoelectrons off of neighboring atoms.  To enable accurate comparisons between compounds, $\mu$(E) has been normalized such that the continuum step at the L$_3$ absorption edge is equal to unity for each sample.  Accordingly, the continuum step at the L$_2$ absorption edge has been normalized to half this value\cite{normalization_note}.  This normalization scheme reflects the number of initial core-electron states available for the L$_2$ and L$_3$ absorption processes, since the ratio of occupied 2p$_{1/2}$ and 2p$_{3/2}$ states is 1:2. 

\begin{figure}
\includegraphics{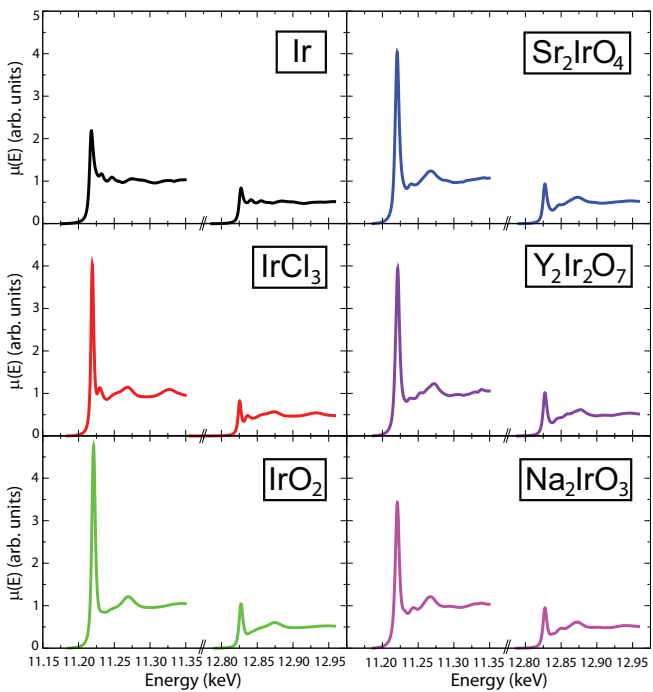}
\caption{(Color online) X-ray absorption spectra collected at the Ir L$_3$ (E = 11.215 keV) and L$_2$ (E = 12.824 keV) absorption edges for a series of Ir-based 5d compounds.}
\end{figure}

The most striking characteristic of the XAS spectra shown in Figure 1 are the prominent white-line features observed at both absorption edges.  The integrated intensity (or area) of these white-line features is proportional to the local density of unoccupied final states in the system (i.e. the population of 5d holes)\cite{Sham_PRB_1985}.  Due to the selection rules that govern electric dipole transitions, the relative strength of the white-line features at the L$_2$ and L$_3$ edges (also known as the branching ratio) can be used to extract information about the total angular momentum, J, of the available 5d hole states.  As $\Delta$J must be equal to 0 or $\pm$ 1 for an allowed dipole transition, it follows that the L$_2$ edge will be exclusively sensitive to transitions involving 5d$_{3/2}$ holes, while the L$_3$ edge will be sensitive to both 5d$_{3/2}$ and 5d$_{5/2}$.  In the limit of negligible spin-orbit coupling effects, the J = 3/2 and J = 5/2 multiplets will be degenerate and the transition probabilities for L$_2$ and L$_3$ processes will only depend on the density of initial core-electron states.  This yields the statistical branching ratio of BR = I$_{L_3}$/I$_{L_2}$ = 2, and an L$_3$ white-line which is twice the size of the L$_2$ feature\cite{vanderLaan_PRL_1988, Thole_PRB_1988, Thole_PRA_1988}.

The presence of strong white-line features indicates that there is a large local density of unoccupied 5d states in all of the Ir compounds measured in this study.  As these white-line features are more pronounced at the L$_3$ edge than the L$_2$ edge, we can also infer that these unoccupied 5d states are primarily 5d$_{5/2}$ rather than 5d$_{3/2}$ in nature.  Note that the ratio of white-line intensities is significantly smaller for pure Iridium than for any of the Ir$^{3+}$ or Ir$^{4+}$ compounds.  The absolute intensity of the L$_3$ white-line is noticeably higher in the Ir$^{3+}$ and Ir$^{4+}$ compounds as well.  Further from the absorption edge, differences in local chemical structure can be observed in the shape of the fine structure oscillations displayed by each compound. 

Similar x-ray absorption spectra are shown for a series of Re-based compounds in Figure 2.  As in the spectra from the Ir-based compounds in Figure 1, each of these samples displays a sharp white-line feature at the L$_2$ and L$_3$ absorption edges.  The integrated intensity of the white-line features grows steadily with increasing oxidation state, as is expected given the rising number of 5d holes in the system.  The ratio of the white-line intensities at the L$_3$ and L$_2$ edges appears to be considerably smaller than that of the Ir-based samples, and much closer to the statistical ratio expected in the absence of spin-orbit coupling effects.  Furthermore, this ratio appears to be roughly equal for each of the Re-based compounds studied, in spite of the fact that the magnitude of the individual white-line features is changing quite significantly.

\begin{figure}
\includegraphics{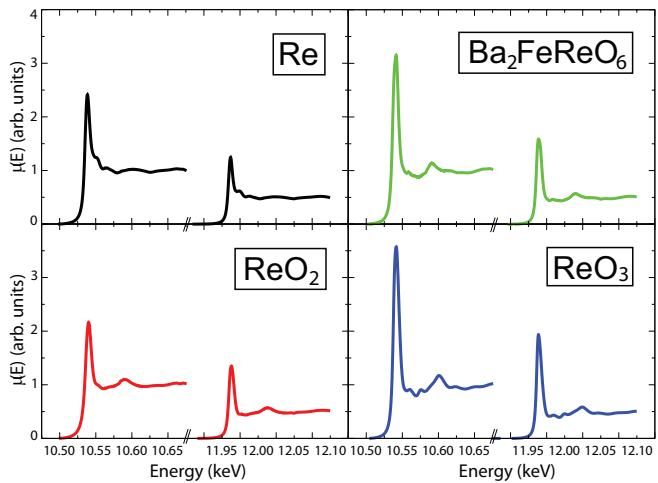}
\caption{(Color online) X-ray absorption spectra collected at the Re L$_3$ (E = 10.535 keV) and L$_2$ (E = 11.959 keV) absorption edges for a series of Re-based 5d compounds.}
\end{figure}

Several systematic effects can be observed by comparing spectra obtained for compounds with different valence states, such as Ir (Ir$^0$, 5d$^7$), IrCl$_3$ (Ir$^{3+}$, 5d$^6$), and IrO$_2$ (Ir$^{4+}$, 5d$^5$).  Representative XAS measurements for these three samples are provided in Figure 3(a).  Clearly, as the ionization state increases there are three significant changes in the XAS spectra at the L$_3$ edge: (1) The position of the white-line feature shifts towards higher energies.  As described in Table II, the fitted peak center shifts from 11218.1(1) eV to 11219.6(2) eV to 11220.7(1) eV following the progression from 5d$^7$ to 5d$^5$.  This is a shift of $\sim$ 1.3 eV for each additional 5d hole.  (2) The integrated intensity of the white-line feature monotonically increases, with fitted peak areas growing by a factor of $\sim$ 2 over the jump from 5d$^7$ to 5d$^5$.  As discussed above, this effect can be understood in terms of the creation of 5d holes, which increase the number of accessible final states for 2p $\rightarrow$ 5d electronic transitions.  (3) The width of the white-line feature varies between oxidation states.  This effect is not monotonic, but the FWHM of the white-line feature varies by $\sim$ 30-50\% between different Ir valence states.  The white-line is sharpest in IrCl$_3$ (FWHM = 3.9(1) eV) and broadest in elemental Ir (FWHM = 6.6(2) eV).  This linewidth is related to the intrinsic bandwidth of the compound, but it is also influenced by broadening mechanisms such as the instrumental resolution, core-hole lifetime effects, and final-state lifetime effects\cite{Leapman_PRB_1982, Sham_PRB_1985}.  These three observations are consistent with previous work by Choy et al, who performed a systematic study of Ir L$_3$ lineshapes in a series of Ir-based double perovskite compounds\cite{Choy_JACS_1995}.  They can also be observed in our Re XAS spectra, following the progression from elemental Re (Re$^0$, 5d$^5$) to ReO$_3$ (Re$^{6+}$, 5d$^1$).

\begin{figure}
\includegraphics{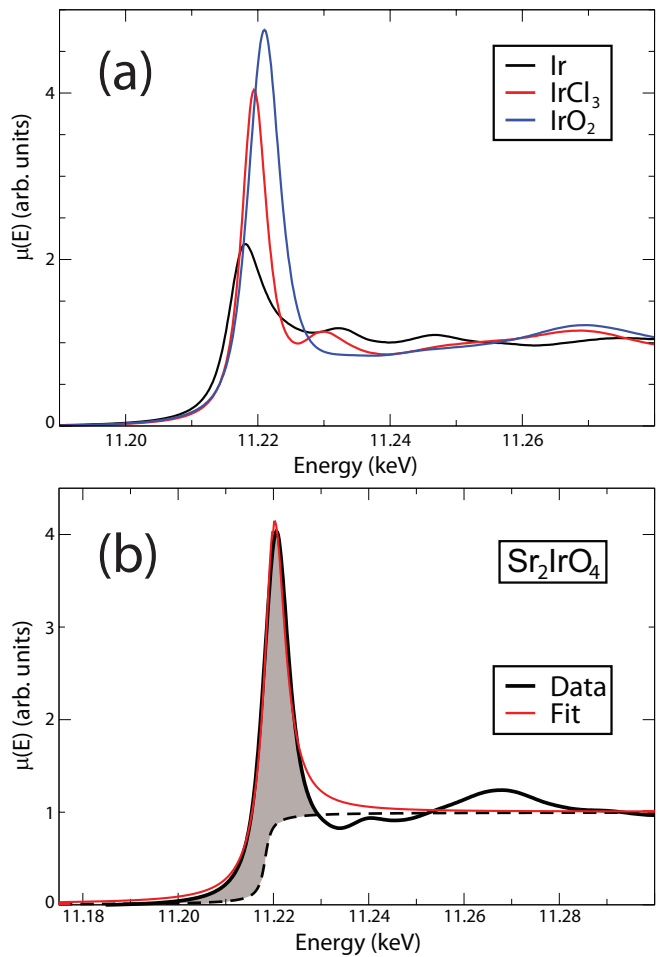}
\caption{(Color online) (a) A comparison of XAS specta at the Ir L$_3$ edge for Ir, IrCl$_3$, and IrO$_2$.  (b) Experimental determination of the white-line intensity at the Ir L$_3$ absorption edge in Sr$_2$IrO$_4$.  The solid black line represents the normalized x-ray attenuation coefficient, $\mu$(E), measured by XAS, while the dashed line represents the arctangent function used to model the continuum step at the L$_3$ absorption edge.  The solid red line represents the best fit to the data using a Lorentzian + arctangent fit function.}
\end{figure}

\begin{table*}
\begin{tabular}{| c | m{1.7cm} | m{0.8cm} | m{1.7cm} |  m{1.7cm} | m{1.8cm} |  m{1.3cm} |  m{1.7cm} | m{1.7cm} | c |}
    \hline
    Compound & \centering{Electronic State} & \centering$\langle$n$_h$$\rangle$ & \centering{I$_{L_3}$ (Numeric)} & \centering{I$_{L_2}$ (Numeric)} & \centering{E$_{L_3}$ [eV]\\ (Fit)} & \centering{$\Gamma_{L_3}$ [eV] (Fit)} & \centering{Branching Ratio (Numeric)} & \centering{Branching Ratio \\ (Fit)} & {$\langle$L$\cdot$S$\rangle$} [$\hbar^2$]\\ \hline \hline
		Ir & \centering{5d$^7$ (Ir)} & \centering{3} & \centering{14.3(6)} & \centering{4.0(2)} & \centering{11218.1(1)} & \centering{6.6(1)} & \centering{3.6(3)} & \centering{3.2(1)} & -1.0(1) \\ \hline
    IrCl$_3$ & \centering{5d$^6$ (Ir$^{3+}$)} & \centering{4} & \centering{21.3(9)} & \centering{3.0(1)} & \centering{11219.6(2)} & \centering{3.9(1)} & \centering{7.1(6)} & \centering{7.8(5)} & -2.5(1) \\ \hline
    IrO$_2$ & \centering{5d$^5$ (Ir$^{4+}$)} & \centering{5} & \centering{32(1)} & \centering{4.7(2)} & \centering{11220.7(1)} & \centering{4.8(1)} & \centering{6.9(6)} & \centering{6.7(4)} & -3.1(1) \\ \hline
    Sr$_2$IrO$_4$ & \centering{5d$^5$ (Ir$^{4+}$)} & \centering{5} & \centering{30(1)} & \centering{4.4(2)} & \centering{11220.2(1)} & \centering{5.5(1)} & \centering{6.9(6)} & \centering{7.0(4)} & -3.1(1) \\ \hline
    Y$_2$Ir$_2$O$_7$ & \centering{5d$^5$ (Ir$^{4+}$)} & \centering{5} & \centering{29(1)} & \centering{5.1(2)} & \centering{11220.0(2)} & \centering{5.5(2)} & \centering{5.8(5)} & \centering{6.0(3)} & -2.8(2) \\ \hline
    Na$_2$IrO$_3$ & \centering{5d$^5$ (Ir$^{4+}$)} & \centering{5} & \centering{25(1)} & \centering{4.5(2)} & \centering{11220.1(2)} & \centering{5.5(2)} & \centering{5.5(4)} & \centering{5.7(3)} & -2.7(2) \\ \hline \hline
    Re & \centering{5d$^5$ (Re)} & \centering{5} & \centering{22.8(9)} & \centering{10.3(4)} & \centering{10538.2(1)} & \centering{8.6(1)} & \centering{2.2(2)} & \centering{2.1(1)} & -0.3(3) \\ \hline
    ReO$_2$ & \centering{5d$^3$ (Re$^{4+}$)} & \centering{7} & \centering{19.0(8)} & \centering{10.6(4)} & \centering{10540.1(2)} & \centering{9.0(3)} & \centering{1.8(1)} & \centering{1.8(1)} & 0.5(4) \\ \hline
    Ba$_2$FeReO$_6$ & \centering{5d$^2$ (Re$^{5+}$)} & \centering{8} & \centering{29(1)} & \centering{13.8(6)} & \centering{10540.7(1)} & \centering{7.7(2)} & \centering{2.1(2)} & \centering{2.1(1)} & -0.3(4) \\ \hline
    ReO$_3$ & \centering{5d$^1$ (Re$^{6+}$)} & \centering{9} & \centering{34(1)} & \centering{16.7(7)} & \centering{10541.3(2)} & \centering{7.4(2)} & \centering{2.0(2)} & \centering{2.1(1)} & -0.1(5) \\ \hline
\end{tabular}
\caption{Summary of results obtained from XAS measurements on Ir and Re-based transition metal compounds.  The intensity of the L$_2$ and L$_3$ white-line features (I$_{L_2}$, I$_{L_3}$), the branching ratio (BR = I$_{L_3}$/I$_{L_2}$), and the spin-orbit expectation value ({$\langle$L$\cdot$S$\rangle$}), have been determined by numeric integration.  The branching ratio, the peak position (E$_{L_3}$) and the width ($\Gamma_{L_3}$) of the white-line feature, have been determined from fits to a Lorentzian + arctangent function.}
\end{table*}

In order to discuss the spin-orbit coupling effects in these compounds on a quantitative level, it is necessary to accurately and precisely determine the intensity of the white-line features observed at the L$_2$ and L$_3$ absorption edges.  The primary method used to obtain these white-line intensities is illustrated in Figure 3(b), using a representative data set collected for Sr$_2$IrO$_4$ at the Ir L$_3$ edge.  The experimental data is described by the solid black line.  Here the continuum edge-step is modelled by an arctangent function, as shown by the dashed black line.  This arctangent is centered at the absorption edge (taken as the inflection point on the rising edge of $\mu$(E)) and is defined to have unit height.  This function can be subtracted from the raw data, leaving only the contributions from the white-line feature (which is approximately Lorentzian in shape) and the fine structure oscillations present at higher energies.  The integrated intensity of the white-line is then calculated by numerically integrating the signal over the range of energies where it is positive-valued (i.e. from below the edge to the point where $\mu$(E) first intersects the arctangent function).  This yields the area of the shaded region in Figure 3(b) which is shown in grey.  This approach has the advantage of not being reliant on a particular choice of lineshape or fit function.  This is important because the mixing of 5d and continuum states can result in broadening of the white-line at higher energies, giving rise to an asymmetric Fano-type lineshape rather than a pure Lorentzian\cite{Qi_PRB_1987}.  The error bars for the integrated intensities are determined from the uncertainty in counting statistics ($\sigma$I = $\sqrt{I}$) and the estimated uncertainty in the position of the arctangent function (which is allowed to vary by $\pm$ 1 eV).  To provide a consistency check, each of the spectra was also fit using a simple Lorentzian + arctangent two-component fit function.  The quality of these fits is illustrated by the solid red line in Figure 3(b).  Although the fit manages to capture the essential characteristics of the white-line feature (the position, the width, and the approximate area), it fails to account for the asymmetric lineshape and the fine structure oscillations near the absorption edge.  Within the limits of the experimental uncertainties, both the numeric integration and the fitting analysis produced the same results.

It is worth noting that in previous studies alternative methods have also been used to model the continuum edge step.  In particular, absorption spectra from compounds such as Au or Pt, which have few 5d holes (and hence negligible white-line intensity), can be scaled, translated, and stretched to provide a relatively clean continuum background\cite{Qi_PRB_1987, Jeon_PRB_1989}.  This approach has proven especially useful in cases where the observed white-line features are weak, as in compounds containing elements with nearly full 5d shells.  For materials that exhibit strong white-line features, such as the Ir and Re-based compounds used in this study, we expect the difference between these two methods to have a negligible influence on the calculated branching ratios.

A list of the numerically integrated white-line intensities obtained for each compound can be found in Table II.  The values of I$_{L_2}$ and I$_{L_3}$ have also been plotted graphically in Figure 4(a).  In the case of the four Re-based samples, both I$_{L_2}$ and I$_{L_3}$ increase approximately linearly with the average number of holes in the system, $\langle$n$_h$$\rangle$.  As discussed above, this is consistent with the increasing density of unoccupied 5d states that accompanies the introduction of additional holes.  It is also clear that I$_{L_3}$ $\sim$ 2I$_{L_2}$, as one might expect to find in the absence of spin-orbit coupling effects.  In the case of the Ir-based compounds, I$_{L_3}$ appears to display the same linear dependence on $\langle$n$_h$$\rangle$, while I$_{L_2}$ remains approximately constant.  A particularly interesting comparison can be made between elemental Re and the four Ir (IV) oxides, since all five of these compounds display the same 5d$^5$ valence electron configuration.  The Ir (IV) oxides are clearly distinguished by a simultaneous enhancement of the L$_3$ white-line feature and suppression of the L$_2$ white-line feature.  However, the sum of the two white-line intensities (I$_{L_2}$ + I$_{L_3}$) remains roughly constant for all five of the 5d$^5$ samples.

The values of the experimentally determined branching ratios are also provided in Table II, and are plotted graphically in Figure 4(b).  Here the branching ratio is defined as BR = I$_{L_3}$/I$_{L_2}$, where I$_{L_n}$ is the integrated intensity of the white-line feature observed at a given absorption edge\cite{BR_convention_note}.  Using theoretical arguments proposed by van der Laan and Thole\cite{vanderLaan_PRL_1988, Thole_PRB_1988, Thole_PRA_1988}, we can relate the branching ratio to the expectation value of the spin-orbit operator $\langle$L$\cdot$S$\rangle$: BR = (2-r)/(1+r), where r = $\langle$L$\cdot$S$\rangle$/$\langle$n$_h$$\rangle$ and $\langle$n$_h$$\rangle$ refers to the average number of 5d holes\cite{LS_convention_note}.  The value of $\langle$L$\cdot$S$\rangle$ is expressed in units of $\hbar^2$.  It should be emphasized that the branching ratio is proportional to $\langle$L$\cdot$S$\rangle$ and not the spin-orbit interaction term in the Hamiltonian, $H_{SO}$ = $\zeta_{5d}${\bf L}$\cdot${\bf S}.  While a large $\langle$L$\cdot$S$\rangle$ does imply the presence of strong spin-orbit coupling effects, the converse is not necessarily true.  In the event that the moments in the system are quenched, it is possible to have $\langle$L$\cdot$S$\rangle$ $\sim$ 0 even if significant spin-orbit interactions are present.

\begin{figure}
\includegraphics{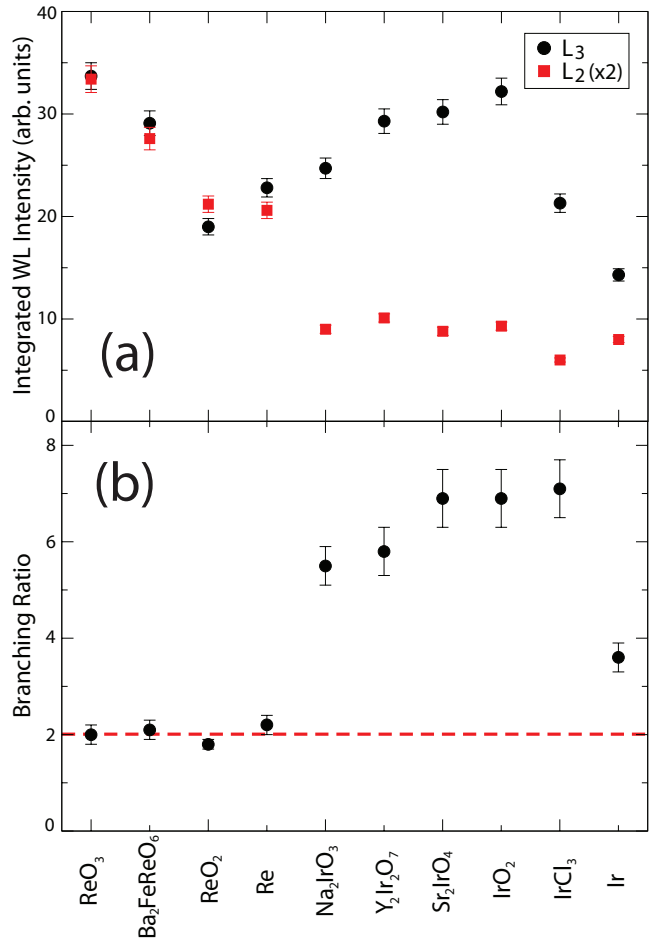}
\caption{(a) Integrated intensity of the L$_2$ and L$_3$ white-line features in Ir and Re-based 5d transition metal compounds.  All intensities were determined by numeric integration, as described in the main text.  (b) Experimentally observed branching ratios, I$_{L_3}$/I$_{L_2}$, determined from the integrated intensities provided in (a).  The dashed line represents the statistical branching ratio of I$_{L_3}$/I$_{L_2}$ = 2 obtained in the limit of negligible spin-orbit coupling.}
\end{figure}

Note that we observe anomalously high branching ratios in all of the Ir-based compounds measured in this study, including IrCl$_3$, IrO$_2$, Na$_2$IrO$_3$, Sr$_2$IrO$_4$, and Y$_2$Ir$_2$O$_7$.  These experimental values are over several times larger than the statistical branching ratio (BR = 2), indicating the presence of extremely large spin-orbit coupling effects.  While this result may not be particularly surprising in the context of Sr$_2$IrO$_4$, Na$_2$IrO$_3$, and Y$_2$Ir$_2$O$_7$ (since each of these samples has previously been associated with unconventional spin-orbit-induced ground states), it is quite unexpected in the case of the IrO$_2$ and IrCl$_3$ reference samples.  Interestingly, the observed branching ratios do not appear to vary significantly between these compounds.  While the ratios observed for Sr$_2$IrO$_4$ and IrO$_2$ appear to be slightly higher, and those for Na$_2$IrO$_3$ and Y$_2$Ir$_2$O$_7$ appear to be slightly lower, these variations do not seem to be correlated with any basic structural properties, such as space group, site symmetry, or local bond length (a full comparison of these properties can be found in Tables I and II).  Perhaps most surprisingly, the branching ratio shows almost no dependence on Ir oxidation state (Ir$^{3+}$ or Ir$^{4+}$) or choice of anion (O or Cl).  This is clearly demonstrated by the IrCl$_3$ data, which is almost indistinguishable from the iridium (IV) oxides.  In contrast, the branching ratios observed for elemental Ir and the four Re-based samples are dramatically different.  Elemental Ir displays a branching ratio of BR = 3.2-3.6, indicating the presence of a moderate spin-orbit enhancement, but significantly less than in any of the other Ir-based compounds studied.  The four Re-based compounds display branching ratios of $\sim$ 2, which is approximately equal to the value of the statistical branching ratio.  This suggests that either the spin-orbit coupling effects in these Re-based transition metal oxides are negligible, or that the quenching of orbital moments results in an extremely small expectation value for the spin-orbit operator L$\cdot$S.  While the latter is certainly a possibility, it would appear to be inconsistent with x-ray magnetic circular dichroism (XMCD) measurements on Ba$_2$FeReO$_6$\cite{Sikora_APL_2006} which indicate that the Re moments are unquenched.

It is interesting to compare these results to the previously reported branching ratios in the literature.  As noted in the introduction, recent XAS measurements by Laguna-Marco and collaborators\cite{Laguna-Marco_PRL_2010} have revealed an anomalously large branching ratio in the layered iridate Ba$_{1-x}$Sr$_x$IrO$_3$ (x = 0, 0.06, 0.12).  The branching ratio quoted in this study is BR $\sim$ 4, a value which is somewhat smaller than the ratios reported in this study, but is still equal to twice the statistical branching ratio.  Similar results have also been reported by Boseggia et al\cite{Boseggia_arXiv_2012}, who performed resonant magnetic x-ray scattering measurements on the bilayer iridate Sr$_3$Ir$_2$O$_7$.  The XAS measurements in this study suggest a BR $\sim$ 5, in very good agreement with the results reported here.  Previous measurements on elemental Iridium also reveal slightly enhanced branching ratios, ranging from BR $\sim$ 2.4\cite{Jeon_PRB_1989} to $\sim$ 4.0\cite{Qi_PRB_1987}.  Our value of BR = 3.2-3.6 falls well within this range.  Somewhat surprisingly, XAS measurements on binary iridium compounds (IrAl, IrAl$_3$, IrSi, and IrSi$_3$)\cite{Jeon_PRB_1989}, iridium alloys\cite{Krishnamurthy_HI_2001, Schutz_ZPB_1989}, and Ir-Fe multilayers\cite{Wilhelm_PRL_2001} are all consistent with approximately statistical branching ratios (BR $\sim$ 2).  This suggests that there must be some property unique to the iridium oxides and chlorides that results in a natural tendency towards large spin-orbit coupling effects.  The observation of enhanced branching ratios in late 5d transition metal oxides such as IrO$_2$, PtO$_2$, and Au$_2$O$_3$, has previously been attributed to a combination of charge transfer and crystal electric field effects\cite{Cho_JPCM_2012}.  As in the case of our Re-based samples, XAS measurements performed on early 5d transition metals (Os, W, Ta)\cite{Qi_PRB_1987, Cho_JPCM_2012} and their associated oxides (OsO$_2$, WO$_2$, and Ta$_2$O$_5$)\cite{Cho_JPCM_2012} all yield branching ratios of $\sim$ 2.  A similar trend is observed in compounds containing lighter 4d elements, such as Ru (RuO$_2$, Sr$_2$RuO$_4$, Sr$_4$Ru$_2$O$_9$, Ca$_{1-x}$Sr$_x$RuO$_3$)\cite{Hu_PRB_2000, Okamoto_PRB_2007} and Rh (Ca$_3$CoRhO$_6$, Ca$_3$FeRhO$_6$)\cite{Burnus_PRB_2008}, which consistently reveal small, statistical branching ratios.

The value of $\langle$L$\cdot$S$\rangle$ can also be compared to previously reported spin-orbit interaction strengths in the literature.  Applying a combination of XAS measurements and configuration interaction calculations, Laguna-Marco et al determined that the expectation value of $\langle$L$\cdot$S$\rangle$ $\sim$ 2 in Ba$_{1-x}$Sr$_x$IrO$_3$ corresponds to a spin-orbit coupling coefficient of $\zeta_{5d}$ $\sim$ 0.3 eV\cite{Laguna-Marco_PRL_2010}.  Although such calculations are beyond the scope of the present study, we note that the coupling coefficient for Sr$_2$IrO$_4$ has been determined to be $\zeta_{5d}$ $\sim$ 0.5 eV from optical conductivity measurements\cite{Kim_PRL_2008}.  Given the expectation value of $\langle$L$\cdot$S$\rangle$ = 3.1 $\pm$ 0.1 determined for Sr$_2$IrO$_4$ from our Ir XAS measurements, there appears to be a relatively simple scaling relation between $\zeta_{5d}$ and $\langle$L$\cdot$S$\rangle$ (i.e. $\zeta_{5d}$ $\sim$ 0.15$\langle$L$\cdot$S$\rangle$).  Based on this relation, we predict $\zeta_{5d}$ will be $\sim$ 0.5 eV for IrO$_2$, and $\sim$ 0.4 eV for IrCl$_3$, Na$_2$IrO$_3$ and Y$_2$Ir$_2$O$_7$.

These results are also relevant to the j$_{eff}$ = 1/2 model which has been put forward to explain the insulating properties of iridates such as Sr$_2$IrO$_4$ and Na$_2$IrO$_3$\cite{Kim_PRL_2008, Kim_Science_2009}.  In the j$_{eff}$ = 1/2 picture, a combination of strong 5d spin-orbit coupling effects and a large octahedral crystal field acts to split the degeneracy of the 5d electronic orbitals, creating a low-lying j$_{eff}$ = 3/2 multiplet (with 4-fold degeneracy) and a higher j$_{eff}$ = 1/2 multiplet (with 2-fold degeneracy).  This process can be regarded from two alternative (although ultimately equivalent) viewpoints: 

\noindent(1) As arising from the action of 5d spin-orbit coupling on the lower t$_{2g}$ and upper e$_{g}$ manifolds created by strong crystal field splitting.  In this case, the j$_{eff}$ = 1/2 and 3/2 states are both derived from the lower-lying t$_{2g}$ levels.  

\noindent(2) As arising from the action of the octahedral crystal field on the lower J = 3/2 and upper J = 5/2 manifolds created by strong 5d spin-orbit coupling effects.  In this case, the j$_{eff}$ = 3/2 states are derived from the lower-lying J = 3/2 levels and the j$_{eff}$ = 1/2 states are split off from the upper J = 5/2 levels.

For an Ir$^{4+}$ ion with a 5d$^5$ electronic configuration (as in Sr$_2$IrO$_4$ and Na$_2$IrO$_3$), the lowest-lying j$_{eff}$ = 3/2 states will be fully occupied, and the j$_{eff}$ = 1/2 band will be partially occupied by a single electron.  At this point, even relatively weak 5d electronic correlations can split the narrow j$_{eff}$ = 1/2 band into an upper Hubbard band and a lower Hubbard band, leading to the development of an insulating gap\cite{Kim_PRL_2008}.  One of the chief methods of distinguishing the j$_{eff}$ = 1/2 ground state from a conventional low spin S = 1/2 state (as produced by large crystal field effects in the absence of strong spin-orbit coupling) is by comparing the ratio of the transition probabilities at the L$_2$ and L$_3$ edges\cite{Kim_Science_2009}.  In the S = 1/2 scenario, the lowest unoccupied state will be found in the t$_{2g}$ manifold, which possesses mixed J = 3/2 and J = 5/2 character.  As a result, both L$_2$ (2p$_{1/2}$ $\rightarrow$ 5d$_{3/2}$) and L$_3$ (2p$_{3/2}$ $\rightarrow$ 5d$_{3/2,5/2}$) transitions will be allowed processes.  In the j$_{eff}$ = 1/2 scenario, the j$_{eff}$ = 3/2 band derived from the J = 3/2 states will be completely occupied, effectively prohibiting any transitions at the L$_2$ edge.  This dramatic difference in transition probabilities should be reflected in the relative strength of the L$_2$ and L$_3$ white-line features, giving rise to the type of strongly enhanced branching ratios observed in this study.

The first experimental signature of the j$_{eff}$ = 1/2 ground state was reported by Kim et al, who observed the characteristic difference in L$_2$ and L$_3$ transition probabilities using resonant magnetic x-ray scattering measurements on Sr$_2$IrO$_4$\cite{Kim_Science_2009}.  The interpretation of these results was subsequently called into question by Chapon and Lovesey\cite{Chapon_JPCM_2011}, who proposed an alternative Ir ground state wavefunction involving a superposition of conjugate states in a Kramers doublet.  This alternative wavefunction was calculated to have a relatively small branching ratio of BR = 1.74\cite{Chapon_JPCM_2011}.  The results of the present study, which determine the branching ratio of Sr$_2$IrO$_4$ to be 6.9 $\pm$ 0.6, are clearly inconsistent with this theoretical prediction.

\section{Conclusions}

In conclusion, we have used x-ray absorption spectroscopy techniques to investigate spin-orbit coupling effects in a series of Ir and Re-based 5d compounds.  We observe anomalously large L$_3$/L$_2$ branching ratios in all Ir-based compounds measured in this study, up to several times larger than the statistical branching ratio of I$_{L_3}$/I$_{L_2}$ = 2.  These enhanced branching ratios indicate that the expectation value for the spin-orbit operator, $\langle$L$\cdot$S$\rangle$, is very large in these systems, reflecting the presence of strong 5d spin-orbit coupling effects.  This observation is consistent with recent proposals for novel spin-orbit-induced physics in iridates such as Sr$_2$IrO$_4$\cite{Kim_PRL_2008, Kim_Science_2009, Jackeli_PRL_2009}, Na$_2$IrO$_3$\cite{Shitade_PRL_2009, Chaloupka_PRL_2010}, and Y$_2$Ir$_2$O$_7$\cite{Pesin_NP_2010, Yang_PRB_2010, Wan_PRB_2011, Witczak-Krempa_PRB_2012}.  Surprisingly, the size of the observed branching ratios does not appear to be sensitive to the detailed structural or electronic properties of these systems.  The sole exception to this rule appears to be provided by elemental iridium, which exhibits a much more modest spin-orbit enhancement, and a branching ratio significantly closer to the statistical value.  The one common feature shared by all of these compounds (save for Ir itself) appears to be the presence of an octahedral crystal field environment.  These anomalously large branching ratios are not common to all 5d compounds, as demonstrated by XAS measurements on a series of Re-based transition metal oxides.  Measurements on these Re-based samples reveal small, statistical branching ratios, a result which implies the presence of minimal spin-orbit coupling effects.

On the basis of these measurements, we propose that even the physics of simple iridium compounds, such as IrO$_2$ and IrCl$_3$, may warrant further investigation due to the presence of unusually large spin-orbit coupling effects.  It may also be productive to extend these studies to include Ir-based compounds which exhibit higher oxidation states (such as Ir$^{5+}$ or Ir$^{6+}$), or which display tetrahedral rather than octahedral local coordination.  In this regard, there appears to be a particularly strong case for complementary studies involving XMCD measurements, similar to those reported in Reference 29.  We hope these results will help to guide and inform future studies of these novel electronic and magnetic systems.

\begin{acknowledgments}

We would like to acknowledge useful discussions with Hlynur Gretarsson, and assistance in sample preparation from Andreea Lupascu and Zixin Nie.  Elements of the data reduction and analysis for this experiment were performed using the Athena software package created by Bruce Ravel.  Work at the University of Toronto was supported by NSERC of Canada, the Banting Postdoctoral Fellowship program, and the Canada Research Chair program.  Use of the HXMA beamline at the CLS is supported by NSERC of Canada, NRC of Canada, CIHR, and the University of Saskatchewan.

\end{acknowledgments}


\begin{thebibliography}{}
\bibitem{normalization_note}
It should be noted that this normalization results in relatively minor changes to the experimental data.  The average correction to $\mu$(E) at the L$_2$ edge is on the order of 5-20\%.
\end{thebibliography}
\end{document}